\documentclass[onecolumn,aps,prb,showpacs]{revtex4}
\usepackage{amsmath} 
\usepackage{amssymb} 
\usepackage{graphicx} 
\usepackage{epstopdf}
\usepackage{color} 
\usepackage{rotating}
 
\begin{document}
 \title{Effective Hamiltonian for the electronic properties of the
quasi-one-dimensional material Li$_{0.9}$Mo$_6$O$_{17}$
}

\author{Jaime Merino}
\email{email:jaime.merino@uam.es}

\affiliation{Departamento de F\'isica Te\'orica de la Materia Condensada and Instituto Nicol\'as Cabrera, 
Universidad Aut\'onoma de Madrid, Madrid 28049, Spain} 

\author{Ross H. McKenzie}
\email{email: r.mckenzie@uq.edu.au}

\affiliation{School of Mathematics and Physics, University of Queensland,
  Brisbane, 4072 Queensland, Australia} 
\date{\today}
                   
\begin{abstract}
 The title material has a quasi-one-dimensional electronic structure and is of considerable interest because
it has a metallic phase
 with properties different from a simple Fermi liquid, a poorly understood "insulating" phase, and a superconducting phase
which may involve spin triplet Cooper pairs. 
Using    the Slater-Koster approach and comparison
with published band structure calculations we present the simplest possible tight-binding model
for the electronic band structure near the Fermi energy. This describes a set of ladders with weak (and frustrated) inter-ladder hopping. In the corresponding lattice model the system is actually close to one-quarter filling (i.e., one electron per pair of sites) rather than half-filling, as has often been claimed. We consider the simplest possible effective Hamiltonian that may capture the subtle competition between unconventional superconducting, charge ordered, and non-Fermi liquid metal phases. We argue that this is an extended Hubbard model with long-range Coulomb interactions. Estimates of the relevant values of the parameters in the Hamiltonian are given.  NMR relaxation rate experiments should be performed to clarify the role of charge fluctuations in 
 Li$_{0.9}$Mo$_6$O$_{17}$ associated with the proximity to a Coulomb driven charge ordering  transition. 
\end{abstract}
\maketitle 

\section{Introduction}

The electronic properties of quasi-one-dimensional materials (Q1D) are of particularly interest because of the possibility that they may exhibit properties characteristic of Luttinger liquid rather the conventional Fermi liquid metallic
state seen in three dimensional metals.
These systems may display Luttinger liquid (LL) behavior above a temperature at which the  thermal energy $k_B T$ is larger than the interchain hopping energy $t_\perp$ \cite{Giamarchi}.
  In a LL charge and spin are carried by independent 
collective excitations instead
of the spin-1/2 charged quasiparticles present in a Fermi liquid. Photoemission experiments
can probe the existence of a LL through the observation of low energy spinon and holon branches    
and a power-law suppression of the density-of-states at the Fermi energy.
Possible realizations of a LL have been investigated extensively in recent years.  Examples are the Q1D cuprate materials,
SrCuO$_2$
 (Ref. \onlinecite{Kim}) 
and Sr$_2$CuO$_3$
 (Ref. \onlinecite{Kidd}),  
 quasi-1D organic crystals \cite{Claessen}, 
carbon nanotubes\cite{Bockrath,Ishii}, GaAs channels \cite{Auslaender,Jompol} and more recently  
one-dimensional Au chains deposited on Ge(001) surfaces\cite{Blumenstein}. 

Apart from the intrinsic interest in finding a LL in actual materials,  quasi-one-dimensional 
materials often exhibit low temperature broken symmetry states including superconducting, 
charge and/or spin density waves and Peierls phases. These instabilities of the metallic phase 
occur at a crossover temperature scale at which the system effectively becomes three-dimensional
and so long range order can occur at low but finite temperatures.
The physics associated with this crossover which goes from the pure one-dimensional to the 
three-dimensional system as temperature is reduced and understanding the possible competing
ground states at low temperatures remains a formidable theoretical challenge.

Another  example of a material
which may exhibit one-dimensional
 physics is the blue bronze  Li$_{0.9}$MoO$_6$O$_{17}$ 
which shows some behavior consistent with a LL \cite{WangPRL06} at sufficiently high temperatures. Angular Resolved 
Photoemission Spectra (ARPES)  shows that quasiparticles do not exist in the system since a 
power law suppression in the DOS is found at the Fermi energy and
dispersing spinon and holon branches 
are seen in the spectral density.  
At sufficiently low temperatures there is a transition to an insulating-like 
or semiconducting state from which superconductivity occurs at even lower temperatures. 
The low temperature "insulating" 
state is poorly understood. The experimental data pose challenging questions to address
apart from the LL behavior observed above the crossover temperature scale. 
In order to understand the mechanism of superconductivity,
it is important to determine the symmetry of the Cooper pairs. It is also important to 
characterize the "insulating"  phase seen below $T_m \sim 25$ K. Finally, it is necessary to characterize 
the nature of the excitations in the metallic state, in particular, if the material is a "bad metal" with 
incoherent excitations or not.  

\begin{figure}
\includegraphics[clip=,width=9cm]{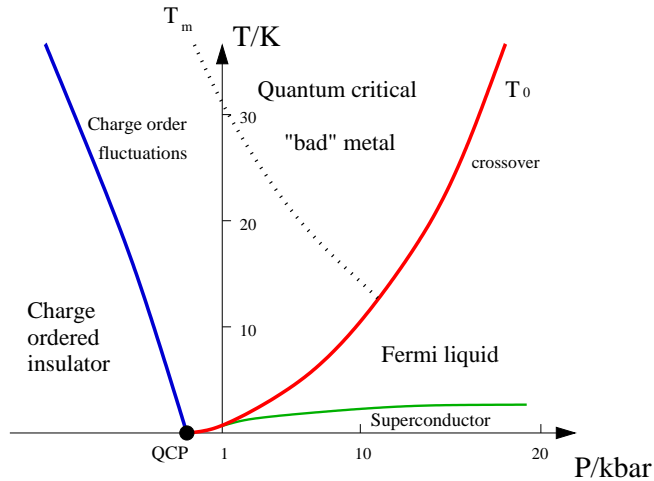}
\caption{(Color online) A speculative temperature-pressure phase diagram relevant to Li$_{0.9}$Mo$_6$O$_{17}$. It assumes
that at ambient pressure the system is close to a quantum critical point for an insulating charge ordered phase. Fluctuations associated
with this phase cause the resistivity to increase with decreasing temperature below T$_m$. The pressure dependence of this curve
is taken from Fig. 3 (a) in Ref. \onlinecite{FilippiniPC89}. Associated with the quantum critical point (QCP) there is a crossover from a quantum critical 
metal (or "bad" metal) to a Fermi liquid metal at a temperature scale $T_0$. The associated coherence temperature goes to zero as the QCP is approached. Similar
physics has been observed in recent studies of a two-dimensional extended Hubbard model at one-quarter filling (cf. Fig. 3 in Ref \onlinecite{Canocortes}). }
\label{fig:phased}
\end{figure}

Here we introduce a realistic microscopic model on a lattice which can capture the essential physics observed. 
This model consists of weakly coupled ladders describing the $d$-orbitals in the Mo atoms with strong on-site 
and off-site Coulomb repulsion. Based on this model we establish that Li$_{0.9}$MoO$_6$O$_{17}$ should 
be regarded as a nearly quarter-filled system and not as a nearly half-filled system. This means that in the absence of longer range Coulomb
interactions the systems would always be a metal. Thus    the long range
part of the Coulomb repulsion should not be neglected and can lead to charge ordering phenomena and a Wigner-Mott insulator.  
The model contains rich physics since it contains low dimensionality,  strong Coulomb interactions, 
geometrical frustration, and charge order frustration. We discuss possible ground states of the proposed model
in different 
limits.  From the model proposed, experimental observations and theoretical considerations we 
speculate that  Li$_{0.9}$MoO$_6$O$_{17}$ is close to a quantum critical point dominated by charge 
ordering fluctuations. A possible
schematic phase diagram is shown in Fig. \ref{fig:phased}.

 The present paper is organized as follows. In Sec. \ref{sec:expt}  we briefly revise key experimental observations
 in Li$_{0.9}$MoO$_6$O$_{17}$.  In Sec. \ref{sec:tb} we discuss the one-electron tight-binding parameters
appropriate for  Li$_{0.9}$MoO$_6$O$_{17}$ based on a Slater-Koster parameterization and comparison with
DFT calculations of the band structure. Sec. \ref{sec:mb} introduces the Coulomb repulsion energies
which are relevant for the description of the title material. This leads to a minimal strongly correlated model which 
consists on weakly coupled ladders comprising the full three dimensional crystal structure. Finally, in 
Sec. \ref{sec:theory} we discuss the theoretical implications of the model obtained and possible consequences.

\section{Brief Review of experiments}  
 \label{sec:expt}
  
We now review some key observed properties
of Li$_{0.9}$Mo$_6$O$_{17}$, with a particular emphasis
on deviations from the behavior seen in conventional metals
and charge ordered insulators. 

\subsection{Electronic anisotropies}

Recent measurements of the resistivity along the three crystal axes yield a
resistivity anisotropy of: $\rho_a : \rho_b : \rho_c \simeq$ 80:1:1600 
at $T=300$ K and $\rho_a : \rho_b : \rho_c \simeq$ 
150: 1: 1600 at $T=4.2$ K. \cite{XuPRL09,WakehamNC11,Mercure}.
Much smaller anisotropies have been observed by other
 authors\cite{LuzPRB07,ChenEPL10}. However,
care must be taken in measuring the conductivity parallel
to the chains because it can be reduced significantly
if the current path is not strictly parallel to the chains.\cite{WakehamNC11}
These are consistent with observed anisotropies in the superconducting upper 
critical fields \cite{Mercure}. Hence, these measurements corroborate the
quasi-one-dimensionality of the system in its electronic properties. 

\subsection{The bad metal}

Transport properties of a wide range of strongly correlated electron materials exhibit certain features that are distinctly different from the electronic properties of elemental metals \cite{MerinoPRB00}. These unusual properties arise from emergence of a low energy scale which defines a temperature scale $T_0$ (often in the range 10-100 K) above which quasi-particles do not exist and the material is referred to as a bad metal. Signatures of this crossover from a Fermi liquid at low temperature to a bad metal at temperatures above $T_0$ include:
(i) the resistivity, Hall coefficient, and thermopower have a non-monotonic temperature dependence, (ii) with increasing temperature the intralayer resistivity smoothly increases to values much larger than the Mott-Ioffe-Regel limit ($h^2 a/e \sim $  1 m$\Omega$cm) corresponding to a mean-free path comparable to the lattice constant $a$ \cite{Gunnarsson}, (iii)	at temperatures of order $T_0$
 the thermopower has values as large as $k_B/e \sim 80 \mu$V/K, and (iv) 
above $T_0$ the Drude peak in the frequency dependent conductivity collapses and the associated spectral weight shifts to higher frequencies.

It is important to realise that if quasi-particles do not exist then the one-electron spectral function does not have dispersive features, i.e., it is completely incoherent. This means that above $T_0$ the notion of a band structure and a Fermi surface have no meaning.
All of the above features in the transport are captured by a dynamical mean-field theory (DMFT)  treatment of lattice Hamiltonians such as the Hubbard model and Anderson lattice model \cite{MerinoPRB00}. It is generally believed that a small $T_0$  is associated with proximity to a Mott insulator or to a quantum critical point \cite{Canocortes}. 
An alternative picture of bad metals can given in terms of 
hard core bosons \cite{Lindner}.

Li$_{0.9}$Mo$_6$O$_{17}$ exhibits many of the above signatures of 
bad metallic behavior.

{\it Resistivity.}
 The magnitude of the resistivity 
parallel to the chains is about 100 $\mu \Omega$cm at 100 K.\cite{XuPRL09,WakehamNC11}
This is comparable to that observed in other "bad metals" such as
optimally doped cuprate superconductors.\cite{Gunnarsson,Hussey}
For a quasi-one-dimensional Fermi liquid the conductivity can be
written as $\sigma_b = 4e^2 \ell /( a c)\hbar $ where $\ell$ is the mean-free path
along the chains and $a$ and $c$ are the unit cell dimensions perpendicular to the chain direction. This leads to an estimate of $\ell \sim 10 b$ at 100 K.

{\it Thermoelectric power.}
The thermopower is positive in sign and a non monotonic function of temperature, increasing from a value
of approximately 20 $\mu$V/K at low temperatures to a maximum of about 
30 $\mu$V/K at 50 K and then decreasing slowly to a value of about 10 $\mu$V/K
around room temperatures \cite{BoujidaPC88}.  
Recent measurements\cite{Cohn} found the magnitude of the
thermopower exhibited a significant sample dependence, with
values as large as 200 $\mu$V/K, and was
interpreted in terms of bipolar transport in an almost
perfectly compensated metal.
All of the above behavior contrasts to what is observed 
in a simple Fermi liquid in which  the thermopower is 
approximately linear in temperature for
temperatures less than $T_0$ \cite{Behnia04}.
Furthermore, in contrast to experiments 
the band structure suggests that the thermopower should be negative since the
system has an electron-like Fermi surface, with two bands close to one-quarter filling.

{\it Frequency dependent conductivity.}
The optical conductivity at both 10 K and 300 K is relatively flat
as a function of frequencies of the scale of tens of meV \cite{ChoiPRB04}, suggesting the absence of a Drude peak. 
The total spectral weight at low frequencies is quite small, corresponding
to about one charge carrier per 10 unit cells if one integrates up
to about 0.1 eV. 
For electric field polarisations perpendicular to the chain direction
there is a clear energy gap of about 0.4 eV, at both 10 K and 300 K.
Hence, the material has the very unusual property that it
appears to be conducting for transport parallel to the chains and
insulating for transport perpendicular to the chains. 

{\it Hall effect.} 
The transverse conductivity $\sigma_{ab}$ increases
by a factor of about 60 as the temperature
decreases from 300 K to 25 K.\cite{WakehamNC11}
The Hall coefficient increases by several orders of magnitude from 
 $(1-3) \times 10^{-9}$m$^3$/C to about 10$^{-6}$m$^3$/C 
as temperature decreases from 200 K to 2 K. 
\cite{ChenEPL10,Cohn}
This is inconsistent with a simple Fermi liquid metal in which the Hall coefficient should be
independent of temperature.  
The value obtained for Li$_{0.9}$MoO$_6$O$_{17}$ assuming a Drude model, $R_H = -1/n|e|$ is 
$R_H \sim -2 \times 10^{-9}$ m$^3$/C, assuming $n=1.9$ electron charge carriers per unit cell.
This value is consistent with the absolute experimental value at $T=250$ K although with the opposite sign.
\cite{1DHall} 
 The $T$-dependence  of $R_H$ shares some similarities 
with the observed behavior in the organic charge transfer salt (TMTTF)$_2$AsF$_6$.
In this material, the Hall coefficient 
increases  by a factor of about 100, as the temperature
decreases from 200 K to 100 K, at which there is a charge 
ordering transition.\cite{Korin-Hamzic}
The Hall coefficient has been calculated for a set of weakly coupled
Luttinger liquids with umklapp scattering associated with one-half filling \cite{LeonPRB}.
Similar results are expected near one-quarter filling.
There are small power law temperature-dependent
corrections to the high temperature non-interacting value. 
The bandwidth sets the scale for this temperature dependence.
In summary, the Hall effect is inconsistent with both
a quasi-1D Fermi liquid and a Luttinger liquid picture even at the qualitative level. 

{\it Thermal conductivity.}
The Lorenz ratio is about 10-30 times the Fermi liquid value,
implying a gross violation of the Wiedemann-Franz law.\cite{WakehamNC11}
This has been interpreted in terms of a Luttinger liquid picture
with spin-charge separation.
The Hall Lorenz ratio increases from about 100 to 10$^5$ times
the Fermi liquid value as the temperature
decreases from 300 K to 25 K.\cite{WakehamNC11}

{\it Angle-Resolved Photoemission Spectroscopy (ARPES).}
This measures the one-electron spectral density
and  results \cite{WangPRL06,WangPRL09} have been interpreted in terms of Luttinger liquid and non-Fermi liquid pictures.
The ARPES data displays characteristic holon and spinon branches as well as the
characteristic suppression of the single particle density of states (DOS) of a one-dimensional system signalling the
destruction of quasiparticles and of Fermi liquid behavior.  However, the exponent $\alpha$ characterizing the 
DOS suppression depends on $T$ varying between: $\alpha \approx 0.6$  for $T \leq 200 K$ and 
$\alpha \approx 0.9$  at $T=300 K$  in contrast with the $T$-independent $\alpha$ in a one band Luttinger liquid
and has been attributed to renormalization of $\alpha$ due to interaction of charge neutral critical modes
associated with the two bands crossing the Fermi energy. 
The scaling of the spectral density with temperature: $A({\bf k}, \omega )=T^\eta {\overline A}(vk/T, \omega/T)$,  violates
the one-band LL scaling relation: $\eta=(\alpha-1)$, since experimentally $\eta=0.56$ instead of $\eta < 0$. 
This unconventional quantum critical scaling indicates the presence of quantum fluctuations 
which can mask the pure LL behavior.  Interestingly no warping of the Fermi surface has been observed in ARPES yet. 

{\it Nernst effect.}
Recent measurements have been interpreted in terms of
bipolar transport in an almost perfectly compensated
metal with close to equal numbers of electrons and holes.
\cite{Cohn}
The magnitude of the Nernst signal is four orders of magnitude
larger than the value given by Behnia's simple Fermi liquid expression.\cite{Behnia09}

{\it Magnetoresistance.}
This exhibits a number of unusual properties, for all
temperatures\cite{XuPRL09,ChenEPL10}.
For magnetic fields parallel to the chains there is a small negative
magnetoresistance, for all current directions, suggesting suppression
of the "insulating" state.
For magnetic fields and currents both perpendicular to the layers
(the c-axis direction) there is a huge positive magnetoresistance.
At fields of 10 Telsa it can be as large as 50 and 500-fold  
 at temperatures of 50 K and 3 K, respectively.\cite{ChenEPL10}
This is non-classical as there is no Lorentz force since the magnetic field
and electric current are parallel.
Similarily, unusual behaviour has also been seen in a wide range
of other strongly correlated low-dimensional metals.

\subsection{The "insulating" state}

Some properties suggest the possibility of an insulating or semiconducting-like phase at low temperatures. 
However, it should be stressed that one does not see activated
behaviour (i.e. clear evidence for an energy gap).
It is observed that the resistivity is a decreasing function of temperature from the superconducting transition
temperature $T_c \sim 1 $ K up to about 20 K.
The resistivity  then increases approximately linear with increasing temperature up to room temperature.
The simplest possible explanation would be that there is a metal-insulator transition around 20 K. However, 
this is inconsistent with several experiments we discuss below.  
First, the minimum in the resistivity occurs at 
$T_m$ at different temperatures for different current directions, ranging from $15 K \leq T_{min} \leq 30 K$.
Increasing the pressure to 20 kbar $T_m$ decreases from
 about 30  to 10  K (Figure 1),
and the magnitude of the resistivity decreases significantly
\cite{FilippiniPC89}.
Hence, it is not clear that the ``insulating'' state exists at high
pressures.
We also note that high magnetic fields parallel to the chain
direction reduce the low temperature resistivity, which
can be interpreted as a destruction of the "insulating" state \cite{XuPRL09}.

The fact that the resistivity is a decreasing function of temperature above
the superconducting transition is rather unusual and puzzling since one normally sees a direct transition from a metallic phase to a superconducting phase. However, there are other cases involving
quarter-filled organic charge transfer salts
 where a superconducting state occurs close to a charge ordered insulator 
[see the Table in Ref. \onlinecite{Merino2001} and
the inset of Fig. 2 in Ref. \onlinecite{Nishikawa}].
Other quasi-one dimensional materials exhibit a resistivity with a similar temperature
and pressure dependence similar to that summarised in Figure 1.
For example, Per$_2$M(mnt)$_2$ [M=Pt,Au] is a CDW insulator at ambient pressure but
above 0.5 GPa the resistivity has a non-monotonic temperature dependence.\cite{Almeida}

An important question is whether the "insulating" state is a 
Charge Density Wave (CDW)  driven by a Fermi surface 
instability with a partially gapped Fermi surface. Such CDWs are observed in other quasi-one-dimensional materials such as 
K$_{0.3}$MoO$_3$ 
and the transition-metal dicalchogenide compound 2H-NbSe$_2$ \cite{Gruner}.  
The structural instabilities associated with the CDW can be clearly seen
in x-ray scattering.
However, high resolution x-ray scattering, neutron scattering \cite{LuzPRB11},
 and thermal expansion\cite{SantosPRL07} experiments on Li$_{0.9}$Mo$_6$O$_{17}$
observe no structural instability such
as one would expect to be associated with charge ordering phenomena. 
However,  observing a structural instability driven by Fermi surface nesting 
requires a sufficiently large electron-lattice coupling which may not be present. 
Other Mo compounds such as the blue bronze 
K$_{0.3}$MoO$_3$ 
which is quasi-two-dimensional show jumps in resistivity typical of a more conventional 
CDW accompanied by a Peierls transition.  A possible interpretation of the observations in
 Li$_{0.9}$Mo$_6$O$_{17}$ is that an unconventional electronically-driven CDW occurs 
 which is not detectable with structural analysis data. 
For example, the "insulating" state may be a D-Density Wave (DDW) state,
which has nodes in the energy gap.
Such a state has been proposed as the low-temperature
state of $\alpha$-(BEDT-TTF)$_2$KHg(SCN)$_4$ (Ref. \onlinecite{Maki}) and of
 the pseudogap state in the cuprates \cite{Chakravarty}.
Since in a DDW state there is no modulation of the charge density 
in real space it is hard to detect the
associated symmetry breaking or long range order. 

It is important to determine the possible nature of the magnetic interactions 
if they exist in the "insulating" phase.  The temperature dependence of the  
magnetic susceptibility has a Curie contribution from a small number of magnetic impurities (about 10$^{-4}$ per conduction electron).\cite{Matsuda}
Subtracting the Curie 
 contribution, the remainder is weakly temperature dependent, 
between about 10 K and 200 K, characeristic of
the Pauli paramagnetism of a Fermi liquid metal.
Hence, there is no sign of the energy gap that is normally seen
in CDWs and charge ordered insulators.
Pauli susceptibility data \cite{Mercure} lead to: $\chi(0) \simeq 2.8 \times 10^{-6}$ consistent
with previous data \cite{Matsuda}.
At low temperatures the specific heat capacity has a term that is approximately linear in 
temperature with a coefficient $ \gamma \simeq$ 6 mJ/(mol K$^2$).\cite{Matsuda,Schlenker}.
More recent data find that the specific heat coefficient at temperatures 
 right above the superconducting transition temperature is: $\gamma \simeq$ 1.6 mJ/ (mol K$^2$).  
From the Fermi velocity obtained from band structure calculations \cite{PopovicPRB06} $\hbar v_F = 3.7 eV \AA$,
one finds a bare density of states at the Fermi energy due to the two Mo($d_{xy}$) one-dimensional bands: 
$D(\epsilon_F)=1.9 $ states/(eV cell). This unrenormalized bare DOS is  three times 
smaller than the DOS obtained from the experimental \cite{Mercure} specific 
heat slope: $\gamma \simeq$ 1.6 mJ  /mol K$^2$).
Experimental data lead to a Sommerfeld-Wilson ratio \cite{McKenzie}:
$R \equiv 4 \pi^2 k_B^2 \chi(0)/(3 (g \mu_B)^2 \gamma) \approx 2$, 
indicating substantial electronic correlation effects.  \cite{Mercure}.

The fact that the charge transport properties
(resistivity and Hall coefficient) show "insulating" behaviour
with decreasing temperature while 
the thermodynamic properties (specific heat and magnetic susceptibility)
show Fermi liquid properties is puzzling. One possible
explanation is the latter are associated with low energy spin excitations and not charge excitations.


\subsection{Superconducting state}

The transition temperature
$T_c$ can vary between about 1-2 K depending on sample purity.
Substituting Li ions with K and Na ions led to a reduction in
$T_c$ and a correlation between $T_c$ with the residual resistivity.\cite{Matsuda}
Such a sensitivity is characteristic of unconventional superconductor,
i.e., non s-wave pairing.\cite{Powell}
Increasing the pressure
to 20 kbar $T_c$ increases from about 1.8 to 2.5 K.
\cite{FilippiniPC89}
The upper critical field for magnetic fields parallel to the chains
 may be above the Chandrasekhar-Clogston-Pauli paramagnetic limit 
for a spin singlet superconductor\cite{Mercure}
suggesting the possiblity of a spin triplet state.

\subsection{Isoelectronic materials}

It might be expected that the materials
 A$_{0.9}$Mo$_6$O$_{17}$
with the alkali metals
A=K,Na,Tl have similar properties and much can be learned from comparisons with A=Li.
However, it turns out that these materials have a slightly different 
crystal structure, with separated metal-oxygen layers
 leading to a significantly different electronic structure.\cite{WhangboJACS87}
Specifically, they turn out to have three
partially filled d-block bands and
a quasi-two-dimensional band structure and Fermi surface.
They undergo CDW instabilities due to a hidden Fermi surface nesting.\cite{WhangboScience91} 
For A=K the CDW transition occurs at 120 K, associated
with a structural transition increasing the 
unit cell dimensions four fold, as seen by X-ray scattering
and electron diffraction.\cite{FilippiniPM84}  At the transition
about 50 per cent of the charge carriers are gapped out.
For A=Na the CDW transition occurs at 80 K, and opening of the
energy gap on two of the bands crossing the Fermi energy
has been seen in angle-resolved photoemission spectroscopy (ARPES).\cite{Glans}

\section{Tight binding band structure}
\label{sec:tb}
In order to introduce the simplest strongly correlated model to describe 
the electronic properties of Li$_{0.9}$MoO$_6$O$_{17}$, we first analyze the band structure published earlier using
the extended H\"uckel method\cite{WhangboJACS88} and
the Local Density Approximation (LDA) version of 
Density Functional Theory (DFT) \cite{PopovicPRB06}.

These calculations allow extracting the nominal valence of the compound,  LiMo$_6$O$_{17}$. 
The band structure shows the existence of two $d_{xy}$ bands crossing the Fermi energy and two $d_{xz}/d_{yz}$ bands 
which are filled with two electrons leading to six electrons in the unit cell.  The $d_{xz}$ and $d_{yz}$ bands are
shifted away from the $d_{xy}$ bands and the Fermi energy becoming completly filled due to hybridization with the neighboring 
O atoms. This effectively leads to two isolated $d_{xy}$ bands crossing the Fermi energy which contain two electrons.  Since there are four Mo atoms per 
unit cell, this implies that each Mo chain atom has 1.5 electrons with one electron in the  $d_{xz}/d_{yz}$  orbitals and 
half an electron in the $d_{xy}$ orbitals. This corresponds to a chemical valence for
 LiMo$_6$O$_{17}$  of Li$^{+1}$(Mo$^{4.5+})_2$(Mo'$^{6+})_4$(O$^{2-})_{17}$
where Mo' denote atoms which are not in the chains.  On the other hand, the 
 doped compound: Li$^{+0.9}$(Mo$^{4.55+})_2$(Mo'$^{6+})_4$(O$^{2-})_{17}$,
has the Mo orbital with supressed electron density: Mo($4d^{1.45}$)  so that there 
is less than 0.5 electrons (about 0.45) per d$_ {xy}$ orbital of Mo.

 \subsection{Slater-Koster parameters}
 
 We parameterize the model using the Slater-Koster 
tight-binding approach \cite{SlaterKoster,Harrison}. 
We use      $2p_x, 2p_y,  2p_z$ orbitals
as the minimum basis
 set for describing O and $4d_{xy}$ orbitals for describing Mo.  Using distances obtained from the 
 X-ray crystal structure \cite{Onoda} we analyze the hopping amplitudes between Mo and O atoms and the direct
 hopping amplitudes between the Mo atoms.  In a second stage, we single out the important hopping amplitudes
needed for an effective tight-binding model Hamiltonian describing the $d_{xy}$ orbitals of Mo.
 
 The distances and displacement vectors needed to define the cosines between the various atoms are shown in 
 Table \ref{table1} and \ref{table2} together with estimated hopping amplitudes obtained from the 
 Slater-Koster approach. The hoppings between the Mo and O atoms are given by: 
 \begin{eqnarray}
 t_{p_xd_{xy}}&=&m(\sqrt{3}l^2V_{pd\sigma} + (1-2l^2)V_{pd\pi}) \\ 
 \nonumber
 t_{p_yd_{xy}}&=&l(\sqrt{3}m^2V_{pd\sigma} + (1-2m^2)V_{pd\pi}) \\
 \nonumber 
 t_{p_xd_{xy}}&=&lmn(\sqrt{3}V_{pd\sigma}-2V_{pd\pi}) \\
 \nonumber 
 \end{eqnarray}
 where $l, m, n$ are the direction cosines of the vector from the $p$ orbital to the $d$ orbital
 and the parameters: 
 \begin{eqnarray}
 V_{pd\sigma}&=&-\frac{29.5 \ {\rm eV} }{d_{MoO}^{7/2} }, \\
 \nonumber 
V_{pd\pi}&=&\frac{13.6 \  {\rm eV}  }{d_{MoO}^{7/2} }.  \\
\nonumber  
 \end{eqnarray}
where the distances are in $\AA$ and the parameters are
 appropriate for Mo. 
According to the above equations one finds that only the $t_{p_yd_{xy}}$ hopping amplitude between Mo and O 
atoms in a chain is non-zero
and the corresponding value shown in Table 
\ref{table2}.  Also  the hopping amplitude between different chains
in a ladder through an O atom is zero with the Slater-Koster approach. 
The direct hopping between two $d_{xy} $ orbitals maybe estimated from:
\begin{equation}
t_{d_{xy}d_{xy}}=3l^2m^2V_{dd\sigma}+(l^2+m^2-4l^2m^2)V_{dd\pi}
\end{equation}
with the corresponding parameters:
\begin{eqnarray}
V_{dd\sigma}=-\frac{213.3 \ {\rm eV}}{d_{MoMo}^5} \\
\nonumber 
V_{dd\pi}=\frac{115.2 \ {\rm eV}}{ d_{MoMo}^5} \\
\nonumber 
\end{eqnarray}
where again the distances are all in $\AA$.
The nearest neighbor intra-chain hopping between Mo is then $t_{Mo_1Mo'_4}=V_{dd\pi}$ at a distance $d=3.725 \AA$ 
and between Mo atoms at $d=3.69 \AA$ in different chains: $t_{Mo_1Mo_4}=V_{dd\delta}=0$  [\onlinecite{Harrison}]. The hopping amplitude between Mo atoms on different ladders is non-negligible: 
$t_{Mo_1Mo'_1}=V_{dd\pi}=0.018 eV$.

 \begin{table}
\caption{Position vectors of the four Mo atoms
 \cite{Onoda} inside the unit cell of Li$_{0.9}$Mo$_6$O$_{17}$ in units of the monoclinic 
$a$, $b$ and $c$ unit cell vectors ($a=12.762 \AA$, $b=5.523 \AA$ and $c=9.499 \AA$). The angle between the $a$ and $c$ axes  
of the monoclinic crystal is $\beta=90.61^0$ }
\label{table1}
\begin{tabular}{llll}
Mo$_i$ &  R$_{Mo_i}a$ &   R$ _{Mo_i}b$  & R$_{Mo_i}c$ \\
\hline 
Mo$_1$ & 0.9939 &  0.25   & 0.23356  \\
Mo$_4$  & 0.16635  & 0.25 &  0.9206  \\ 
Mo'$_1$  & 0.00613 &  0.75  & 0.7664 \\
Mo'$_4$  &   0.8337 &  0.75   & 0.07938 \\
\hline
\end{tabular}
\end{table}

\begin{table}
\caption{Distances between molybdenum and oxygen atoms \cite{Onoda} and hopping amplitudes in Li$_{0.9}$Mo$_6$O$_{17}$. 
Hoppings are estimated based on the Slater-Koster approach. The most relevant nearest-neighbor Mo-Mo
and Mo-O distances within the main ladder and between different ladders are displayed. Hopping amplitudes are between $4d_{xy}$ orbitals  
Mo atoms and between $4d_{xy}$ and $2p_y$ orbitals for Mo-O within a chain.  Distances are given in $\AA$ and hopping in eV.
The labeling of the atoms follows Ref. \onlinecite{Onoda} (see Fig. 2).}
\label{table2}
\begin{tabular}{lll}
Atom-atom   &  Distance($\AA$)   & Hopping (eV)  \\
\hline 
Mo$_1$-Mo'$_4$ (in-chain)  &  3.725     &    0.1606  \\
Mo$_1$-O$_{11}$  (in-chain) & 1.873   & 1.515  \\
Mo$_1$-Mo$_4$ (inter-chain) & 3.6756  & 0.  \\ 
Mo$_1$-O$_1$ (inter-chain) & 1.873  &   0. \\
Mo$_1$-Mo'$_1$ (inter-ladder) &   5.7655  &  0.018 \\
\hline
\end{tabular}
\end{table}

The largest contribution to the interaction between neighboring Mo atoms in the chain comes from 
hopping through intermediate O atoms. A straightforward estimate of this hopping amplitude from perturbation theory gives:
\begin{equation}
t(Mo-Mo)={t^2_{p_y d_{xy}} \over \epsilon_{Mo}-\epsilon_O}=0.9 \ {\rm eV}, 
\end{equation}
which leads to a half-bandwidth of $1.8$ eV  for the chain,
a value which is large
 compared to the LDA-DFT value of about $0.9-1$ eV
 (Ref. \onlinecite{PopovicPRB06}).  
Such a discrepancy may be related to the breakdown of the perturbative form since:$ |\epsilon_{Mo}-\epsilon_O|  \sim  t_{p_y d_{xy}}$
using
the values $\epsilon_{Mo}(4d)=-11.56$ eV and $\epsilon_O(2p)=-14.13$ eV.
We will discuss in more detail below 
how the effective nearest-neighbor hopping between Mo atoms needs to be reduced by a factor of two with respect to the Slater-Koster hoppings
in order to correctly reproduce the LDA-DFT bandwidths. 

From the Slater-Koster analysis above we would arrive at the conclusion that the zig-zag Mo chains 
are then decoupled and only a non-zero inter-ladder hopping exists. However,  DFT band structure calculations do indicate 
the existence of inter-chain hopping which is, however, 
much weaker than the effective in-chain hopping. This is because the direct Mo$_1$-Mo'$_4$ bonding 
in the chain is essentially of the $\pi$-type whereas the interchain Mo$_1$-Mo$_4$ bond is of the weak $\delta$-type 
since the chains are on top of each other. Therefore, the inter-chain coupling should come from the small $\delta$-type coupling
between the $d_{xy}$ orbitals of the Mo atoms. 
Ladders are coupled through a direct hopping connecting Mo $d_{xy}$ orbitals. The interladder Mo$_1$-Mo'$_1$  bonding is 
of the $\pi$-type. On the other hand, hopping processes via two intermediate O atoms should be very small.  This is because one of these 
intermediate oxygens is an apical O, (O$_4$ in Fig. 2 of Ref. \onlinecite{Onoda}) weakly hybridized 
 to Mo$_1$ via the O$_4$ and O'$_8$ $p$-orbitals.  It is important to note that interchain and interladder hoppings
should be non-zero in order to recover the dispersion along the $c$-direction found in DFT caculations.
A three parameter Slater-Koster approach \cite{Harrison} to the bands 
would lead to an interchain hopping amplitude: $t_{Mo_1Mo_4}=V_{dd\delta}=V_{dd\sigma}/6=-0.046$ eV which 
has the opposite sign to the $\pi$-type of bond between different ladders $t_{Mo_1Mo'_1}=0.018$ eV. We will see
below how the opposite signs between these two hopping amplitudes are important in a tight-binding model 
to capture the LDA-DFT band structure close to the Fermi energy.

\subsection{Effective tight-binding model for Li$_{0.9}$Mo$_6$O$_{17}$}
We have performed a tight-binding calculation in which we include the four $d_{xy}$ orbitals of the 
Mo atoms and integrate out the $O$ atoms which are assumed to enter indirectly via the $Mo-Mo$ hopping 
amplitudes.  We keep the most relevant hopping amplitudes needed for the description of the 
band structure. The simple tight-binding model for the $d_{xy}$ orbitals
is compared to previous LDA-DFT band structure calculations and the hopping amplitudes extracted. 

 The position of the atoms in the unit cell are expressed as: 
${\bf R}_i= R_{ia}{\bf a}+ R_{ib}{\bf b}+R_{ic}{\bf c}$ and the
momentum wavevector: ${\bf k}=k_a {\bf a^*} + k_b {\bf b^*} + k_c {\bf c^*} $, all refered to the unit cell coordinate system. 
A translation vector of the lattice reads: ${\bf R_n}= n_a {\bf a} + n_b {\bf b} + n_c {\bf c} $ and the Bloch functions:
\begin{equation}
|  \Psi_{Mo_i}({\bf k})  \rangle = {1 \over \sqrt{N} } \sum_{\bf R_n} e^{i {\bf k}\cdot ({\bf R}_n+{\bf R}_{Mo_i})}  | \Psi_{Mo_i}({\bf r-R_n-R_{Mo_i}})  \rangle
\end{equation}

For the effective in-chain interaction we have:

\begin{equation}
\langle    \Psi_{Mo_1}({\bf k})  | H  |  \Psi_{Mo'_4}({\bf k})  \rangle= \langle    \Psi_{Mo_4}({\bf k})  | H  |  \Psi_{Mo'_1}({\bf k})  \rangle= -2tA(k_a,k_c) \cos({{ \bf k}\cdot{\bf b}  \over 2}),
\end{equation}
where $A(k_a,k_c)=e^{i 2\pi (k_a (R _{Mo'_4}-R_{Mo_1})_a+k_c (R _{Mo'_4}-R_{Mo_1})_c) }$.
The interaction between chains in the same ladder is described through matrix elements as:
\begin{equation}
\langle    \Psi_{Mo_1}({\bf k})  | H  |  \Psi_{Mo_4}({\bf k})  \rangle= 
-t_\perp e^{i{\bf k}\cdot ({\bf a} -{\bf c})}e^{i {\bf k}\cdot ({\bf R _{Mo_4}-R_{Mo_1}}) } 
=-t_\perp e^{i{\bf k}\cdot{\bf \delta}_\perp }
\end{equation}
where ${\bf \delta_\perp} = 0.17 {\bf a} -0.31 {\bf c}$,
and similarly for the $Mo'_4$-$Mo'_1$ interaction:
\begin{equation}
\langle    \Psi_{Mo'_1}({\bf k})  | H  |  \Psi_{Mo'_4}({\bf k})  \rangle=-t_\perp e^{-i{\bf k}\cdot{\bf \delta}_\perp } =
\langle    \Psi_{Mo_1}({\bf k})  | H  |  \Psi_{Mo_4}({\bf k})  \rangle^*
\end{equation}
Two  ladders in two neighboring unit cells are coupled through the matrix element:
\begin{eqnarray}
\langle    \Psi_{Mo_1}({\bf k})  | H  |  \Psi_{Mo'_1}({\bf k})  \rangle 
&=&
-t' e^{i{\bf k}\cdot{\bf a}}e^{i{\bf k}\cdot ({\bf R _{Mo'_1}-R_{Mo_1}}) } -t'  e^{i{\bf k}\cdot({\bf a} -{\bf b})} 
e^{i{\bf k}\cdot ({\bf R _{Mo'_1}-R_{Mo_1}}) } \nonumber \\
&=& -2t'e^{i{\bf k}\cdot{\bf \delta_1}}\cos({{ \bf k}\cdot{\bf b} \over 2})
\end{eqnarray}
where ${\bf \delta_1} \equiv {\bf a} - {\bf b}/2 + {\bf R _{Mo'_1}-R_{Mo_1}}
=0.01 {\bf a} + 0.53 {\bf c}$.
On the other hand the $Mo_4-Mo'_4$ hopping amplitude between ladders is weaker since they are at a larger
distance and is neglected:
\begin{equation}
\langle    \Psi_{Mo_4}({\bf k})  | H  |  \Psi_{Mo'_4}({\bf k})  \rangle=0.
\end{equation}

The final tight-binding $4 \times 4$ Hamiltonian to be diagonalized reads:

\[ H(k)= \left(  \begin{array}{cccc}
0 & -t_\perp e^{i{\bf k}\cdot{\bf \delta_\perp }} & -2t'e^{i{\bf k}\cdot{\bf \delta_1}}\cos({ { \bf k}\cdot{\bf b}  \over 2}) & -2tA(k_a,k_c)\cos({ {\bf k}\cdot{\bf b} \over 2})    \\
-t_\perp e^{-i{\bf k}\cdot{\bf \delta_\perp }} &0 & -2tA(k_a,k_c)\cos({{ \bf k}\cdot{\bf b} \over 2})  &  0   \\
-2t'e^{-i{\bf k}\cdot{\bf \delta_1}}\cos({{\bf k }\cdot {\bf b} \over 2})  & -2tA(k_a,k_c)\cos({{\bf k}\cdot{\bf b}  \over 2}) & 0 &  -t_\perp e^{-i{\bf k}\cdot{\bf \delta_\perp }}   \\
-2tA(k_a,k_c)\cos({{\bf k}\cdot{\bf b}  \over 2}) & 0 &  -t_\perp e^{i{\bf k}\cdot{\bf \delta_\perp }}  &  0  \\

\end{array} \right) \]

In the absence of interladder hopping ($t'=0$), the four bands:
\begin{equation}
\epsilon({\bf k}) = \pm t_\perp \pm 2 t  \cos({{ \bf k}\cdot{\bf b} \over 2}),
\end{equation}
which recovers the expected dispersions of  the uncoupled two-leg ladders, as it should.

\begin{figure}
\includegraphics[clip=,width=5cm]{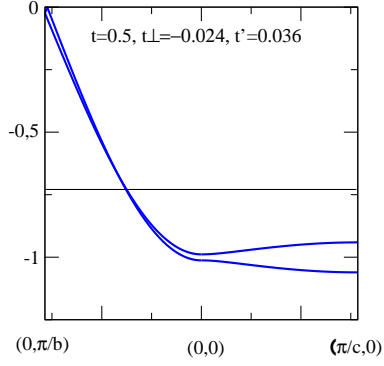}
\caption{(Color online) Band structure for the lower two
bands obtained from the diagonalization of the $4\times4$ 
tight-binding matrix defined in the text using $t=0.5$ eV, $t_\perp=-0.024$ eV, $t'=0.036$ eV.  These parameter values are chosen to produce band dispersions
comparable to LDA-DFT calculations \cite{PopovicPRB06}. The band dispersions along the $b$ and 
$c$ directions are plotted. The horizontal line denotes the Fermi
energy.}
\label{fig:bands}
\end{figure}

Diagonalizing the Hamiltonian we obtain the band structure along the $b$ and
 $c$-directions shown in Fig. \ref{fig:bands}. The effective hopping between the nearest-neighbor  Mo$_1$-Mo'$_4$ atoms 
 in a chain is taken to be $t=0.5$ eV in order to reproduce the DFT bandwidth of the two lowest Mo $d_{xy}$ energy bands. 
We use the hopping amplitudes $t'=0.036$ eV a factor of two larger than the one obtained 
from Slater-Koster in Table \ref{table2} and we take the hopping between chains in the ladder opposite to $t'$ and of magnitude: $t_\perp=-0.024$ eV
consistent with the type of bonding ($\pi$ vs. $\delta$ bonding). 
This set of parameters captures the correct magnitude of the band dispersions 
along $b$, the dispersion of the two lowest bands in the $c$-direction due to the combined coupling between ladders, $t'$,  
and the interchain hopping amplitude,  $t_\perp$.   The magnitude of the values extracted from the present analysis 
$t_\perp$ and $t'$ of about 0.03 eV are consistent with recent preliminar estimates \cite{WangPRL09}.

\begin{figure}
\includegraphics[clip=,width=5cm]{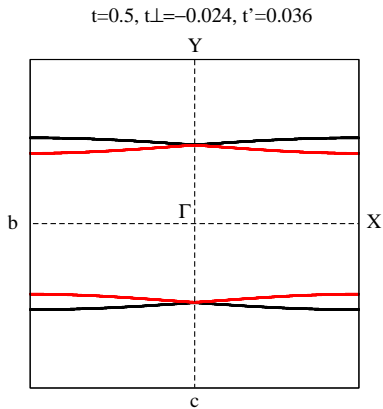}
\caption{(Color online) Fermi surfaces obtained from the diagonalization of the $4\times4$ 
matrix using $t=0.5$ eV, $t_\perp=-0.024$ eV, $t'=0.036$ eV and filling $n=1.9$.
The Fermi surfaces corresponding to the two separate bands are denoted
in blue and red. Note that the two sheets have opposite warping.
}
\label{fig:fermistb}
\end{figure}

The Fermi surface obtained from the diagonalization of the $4 \times 4$ Hamiltonian is shown in Fig. \ref{fig:fermistb}.
This tight-binding Fermi surface is close to the DFT Fermi surface
 (cf. Fig. 6.  in Ref. \onlinecite{PopovicPRB06}) except for the latter the two Fermi surfaces touch at 
the zone boundary.  The filling is $n=1.9$,  with
 the warping of the most filled band opposite to the less filled band. 
 We note that the opposite signs of $t'$ and $t_\perp$ found in the Slater-Koster approach 
 is essential for capturing the opposite warping of the two Fermi sheets found in DFT.
The Fermi surface obtained from the diagonalization of the $4 \times 4$ Hamiltonian is shown in Fig. \ref{fig:fermistb}.

\begin{figure}
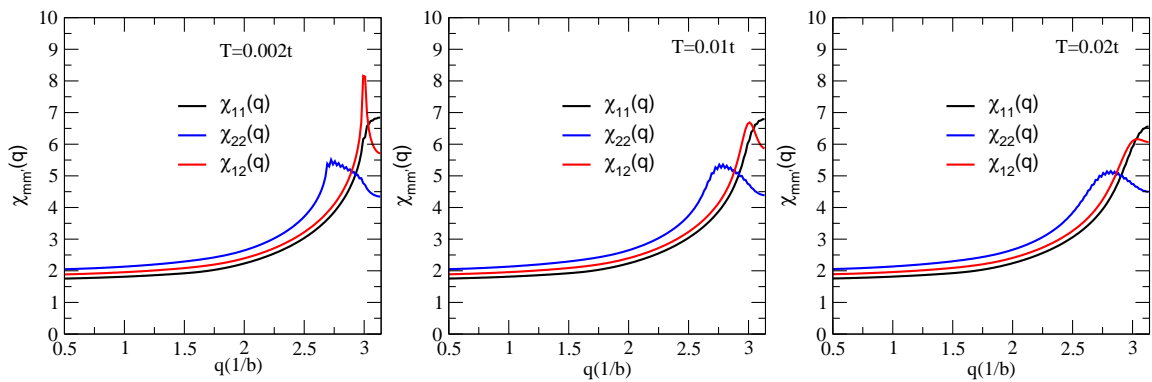

\includegraphics[clip=,width=5cm]{fig4a.eps}
\includegraphics[clip=,width=5cm]{fig4b.eps}
\includegraphics[clip=,width=5cm]{fig4c.eps}
\caption{(Color online) Wavevector dependence of the three
 non-interacting charge susceptibilities.
The three boxes correspond to decreasing temperature from left to right.
The three susceptibilities are associated with
the two low energy bands crossing the Fermi surface and shown 
in Fig. \ref{fig:fermistb} with 
 $t=0.5$ eV, $t_\perp=-0.024$ eV, $t'=0.036$ eV and filling $n=1.9$. The interband susceptibility $\chi_{12}$
 displays an instability at low temperatures due to significant
nesting between these two Fermi sheets. 
}
\label{fig:chi0}
\end{figure}

{\it Fermi surface nesting instabilities.}
The tendency of the system towards nesting instabilities can be analyzed by computing the 
bare charge susceptibility, 
\begin{equation}
\chi_{mm'}({\bf q})={2 \over N}\sum_{\bf k} {f(\epsilon_m({\bf k +q}))-f(\epsilon_{m'}({\bf k})) \over \epsilon_m({\bf k +q})- \epsilon_{m'}({\bf k}) }, 
\end{equation}
between the different  bands, $\epsilon_m({\bf k})$. From a simple
inspection of the Fermi surface in Fig. \ref{fig:fermistb} one observes that since the two Fermi sheets
corresponding to the two bands are oppositely warped the nesting condition is optimally satisfied 
with a nesting vector which is about ${\bf q} \approx (0,\pi/b,0)$ between the two different bands. Evaluating 
 $\chi_{mm'}({\bf q})$ shown in Fig. \ref{fig:chi0}, a divergence is observed in the interband
 susceptibility: $\chi_{12}({\bf q})$ when decreasing the temperature but not in the intraband contributions. 

\subsection{Quarter filling or half filling?}

From a band structure point of view the apparent half filling arises because of the zig-zag structure of the chains
which leads to a folding of the Brillouin zone.
However, the key point (at least in terms of
strong electronic correlations) is that for LiMoO$_6$O$_{17}$
 there would be one
electron per two Mo ions.
In this sense the material should be viewed as being close to one-quarter filling.
This is important because it means that long-range Coulomb interactions
are required to produce an insulating state.
We also note that both ARPES and LDA give  $k_F \simeq \pi/2b$ and this only corresponds
to half filling when $b$ is the nearest-neighbour distance. 

\subsection{Anisotropy in the conductivity }

In a simple Fermi liquid picture
the ratio of the conductivity parallel and perpendicular to the chain
directions is independent of temperature and given by
\begin{equation}
\frac{\sigma_b}{     \sigma_a}
\sim \left(
\frac{t}{t'} \right)^2.
\end{equation}
The estimates in Table \ref{table:parameters}
suggest an anisotropy ratio of about 100 which
is consistent with that observed experimentally
in References \onlinecite{XuPRL09}, \onlinecite{Mercure}, and \onlinecite{WakehamNC11},
but much larger than reported in References \onlinecite{LuzPRB07} and \onlinecite{ChenEPL10}.

\section{Effective strongly correlated Hamiltonian}
\label{sec:mb}

\begin{figure}
\includegraphics[clip=,width=8cm]{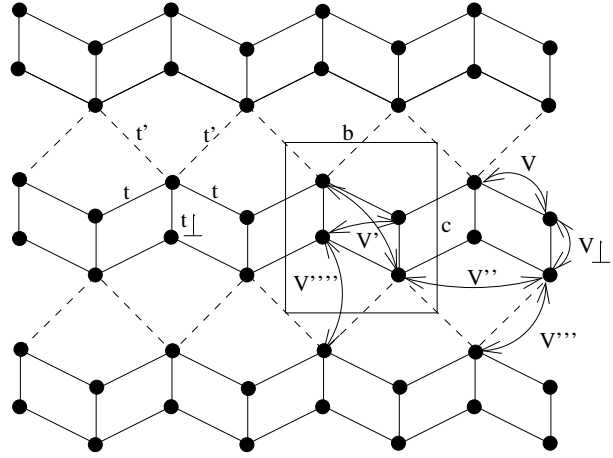}
\caption{(Color online) Schematic representation of the model 
Hamiltonian (\ref{eq:model}) to describe Li$_{0.9}$Mo$_6$O$_{17}$.
The hopping and Coulomb energies entering the model describing electrons in the d$_{xy}$ orbitals of $Mo$ are displayed.
A schematic view perpendicular to the $b-c$ plane is shown noting that the planes of the ladders in the real crystal are tilted 
with respect to the $b-c$ plane. The rectangle corresponds to the unit cell in the b-c plane.
}
\label{fig:struct}
\end{figure}

An effective Hamiltonian which can capture the electronic properties of Li$_{0.9}$MoO$_6$O$_{17}$ is now presented. 
We argue that since the system is close to quarter-filling  not only the onsite $U$ but also the
off-site Coulomb repulsion $V$ should be included to account for charge ordering phenomena.
This minimal model is an extended Hubbard model which reads:
\begin{equation}
H=H_0+H_U,
\label{eq:model}
\end{equation}
where $H_0$ is the non-interacting tight-binding Hamiltonian which reads:
\begin{eqnarray}
H_0&=&-t \sum_{\alpha,l,i\sigma}(c^{(l)\dagger}_{\alpha,i\sigma} c^{(l)}_{\alpha,i+1\sigma} + c.c.) 
-t_\perp \sum_{\alpha,l,i\sigma} (c^{(l)\dagger}_{\alpha,i\sigma} c^{(l+1)}_{\alpha,i\sigma} + c.c.)
\nonumber \\
&=&-t'\sum_{\alpha,i\sigma}(c^{(l)\dagger}_{\alpha,2i-1\sigma} c^{(l+1)}_{\alpha+1,2i\sigma} + c.c)
-t'\sum_{\alpha,i,\sigma}(c^{(l)\dagger}_{\alpha,2i+1\sigma} c^{(l+1)}_{\alpha+1,2i\sigma} + c.c.),
\nonumber \\
\label{eq:tb}
\end{eqnarray} 
where the first term describes the kinetic energy of a single ladder and the second term the 
hopping processes between ladders.  The index $l$ denotes one of the two chains in a ladder, $\alpha$ labels a specific ladder and 
$i$ runs from 1 to $N$, the number of sites in the chains.  
The parameters $t$, $t_\perp$,  denote hopping amplitudes within a ladder whereas 
$t'$ denotes the hopping amplitude connecting nearest-neighbor ladders. One only needs three hopping amplitude parameters
for the description of the band structure.  
The Coulomb interactions are encoded in $H_U$ which reads:
\begin{eqnarray}
H_U &=& U\sum_{\alpha,l,i\sigma} n^{(l)}_{\alpha i \uparrow} n^{(l)}_{\alpha i\downarrow}
+\sum_{\alpha,l,i} V n^{(l)\dagger}_{\alpha,i} n^{(l)}_{\alpha,i+1} +V_\perp \sum_{\alpha,l,i} n^{(l)}_{\alpha,i} n^{(l+1)}_{\alpha,i}+V' \sum_{\alpha,l,i} (n^{(l)}_{\alpha,i} n^{(l+1)}_{\alpha,i+1}+
n^{(l)}_{\alpha,i+1} n^{(l+1)}_{\alpha,i} )+V''\sum_{\alpha,l,i}n^{(l)}_{\alpha,i} n^{(l)}_{\alpha,i+2}
\nonumber \\
&+&V'''\sum_{\alpha,l,i}n^{(l)}_{\alpha,i} n^{(l+1)}_{\alpha+1,i+1}+V''''\sum_{\alpha,l,i}n^{(l)}_{\alpha,i} n^{(l+1)}_{\alpha+1,i}.
\nonumber \\
\label{eq:hu}
\end{eqnarray}
The parameters $U$, $V$, $V_\perp$, $V'$ and $V''$  denote intraladder Coulomb interactions, 
whereas $V'''$, $V''''$ and $V''''$ denote Coulomb interactions between nearest-neighbor 
ladders. A schematic representation of the parameters entering the model proposed are displayed in Fig. \ref{fig:struct}. 
In principle,  the model keeps the essential intra-ladder and inter-ladder Coulomb repulsion energies that 
may be relevant for Li$_{0.9}$MoO$_6$O$_{17}$. 
Estimates of the values of Coulomb repulsion energies entering the Hamiltonian 
are provided and discussed below.

\subsection{Model parameter values}

In Table \ref{table:parameters} we show the hopping amplitudes and Coulomb parameters estimated for model (\ref{eq:model}). 
From the comparison to the DFT-LDA
 calculations we found that the nearest-neighbor hopping is $t=0.5$ eV,  the hopping between 
chains in the same ladder is $t_\perp=-0.024$ eV and between nearest-neighbor
chains in different ladders: $t'=0.036$ eV. We note that for the case in which $t'=-3t_\perp/2$, we have that the two bands cross at $k_c=0$
in agreement with the DFT calculations.  When the same sign is used $t'=3t_\perp/2$, then the two bands do not cross at $k_c=0$.
So it is essential that $t_\perp$ and $t'$ do have opposite signs to capture the appropriate warping and dispersions. 
The on-site Coulomb repulsion is about $U=6.4$ eV estimated from constrained-DFT \cite{PopovicPRB06} calculations.

The unscreened Coulomb interaction between electrons in nearest neighbor Mo atoms within a chain is estimated to be
 $V=2$ eV, which is comparable to the bandwidth and therefore relevant unless complete screening inside 
 the crystal occurs. These values are large enough to lead to charge ordered ground states since the critical 
 value $V_c$  in a quarter-filled chain is about: $V \approx 2t$.  Screened values of $V$ 
 are reduced to about $V=0.2$ eV, assuming a dielectric constant of
$\epsilon \approx 10 \epsilon_0$ due to the screening of the rest of the crystal. 
\cite{PopovicPRB06} 

In a similar quarter-filled ladder \cite{Smolinski} compound NaV$_2$O$_5$ charge ordering phenomena due 
to the off-site Coulomb repulsion has been found. From
DFT-based calculations the parameters are
estimated to be: $t_\parallel \simeq 0.17$ eV,  $t_\perp \simeq 3 t $, and $U \simeq 2.8 $ eV.
The distance between vanadium ions along the chain is approximately
$3.8 \AA$ and the nearest-neighbor Coulomb repulsion $V$ has been estimated\cite{Bernert} to be $V \approx 2t$
consistent with the zig-zag type charge ordering \cite{Noack} found in these systems. 

Experimental observations combined with the present discussion suggest
that in Li$_{0.9}$Mo$_6$O$_{17}$,  it 
is desirable that the Coulomb repulsion 
$V \sim 0.5-1$ eV so that the system is placed near the QCP to a charge order transition driven by $V$ (cf. Fig. \ref{fig:phased}).  
This requires that the long-range Coulomb repulsion is only partially screened and is important in order to understand 
the electronic properties of Li$_{0.9}$Mo$_6$O$_{17}$. 
The relevance of incomplete Coulomb screening to the electronic properties of quasi-one-dimensional systems 
has already been pointed out in the case of TTF-TCNQ\cite{Hubbard,CanoCortes07} crystals.

We now discuss the relevance of further neighbor Coulomb interactions. 
The Coulomb repulsion between chains in the same ladder should be similar to 
the nearest-neighbor $V$, since the 
distance between two neighboring Mo atoms in 
the same chain is 3.725 $ \AA$  and  is 3.675 $\AA$ between
atoms in different chains, so $V_\perp \approx V$. 
This parameter is important for stabilizing a zig-zag type of charge  ordered state in the ladder. 
Comparing the distances between different Mo atoms 
one realises that further neighbor interactions are relevant to the model. In fact, 
the Coulomb repulsion between electrons on Mo atoms on different zig-zag chains of the same ladder (see Fig. \ref{fig:struct}), $V'$ is
comparable to $V$ and $V_\perp$ since interladder
distances are between 5.256 and 5.366 $\AA$. 
Since the next-nearest neighbors distance between two Mo is 5.52 $\AA$ due to the zig-zag nature of the chains,  
we have that $V''$ is not be negligible.  On the other hand the nearest distance between Mo atoms in nearest ladders is 5.77 $\AA$ so the
Coulomb repulsion between neighboring ladders $V'''$ is comparabe to $V''$.  
The off-site Coulomb repulsion included in the model is cut-off at values of the Coulomb repulsion at which: 
$V'''' \sim V/2 \sim t $, assuming $V \sim 2t$. Coulomb energies between 
farther distant neighbors are neglected since they are smaller than the bare kinetic energy of the electrons. 

Hence,  a minimal model should contain the nearest-neighbor Coulomb interactions
and possibly longer range Coulomb interactions between electrons in further distant Mo atoms as represented in Fig. \ref{fig:struct}. 
Accurate screening calculations in Li$_{0.9}$Mo$_6$O$_{17}$ should be performed in order to 
pin down the values of the off-site Coulomb repulsion and settle the relevance of the longer range 
Coulomb repulsion in Li$_{0.9}$Mo$_6$O$_{17}$.  Below we discuss the 
minimal strongly correlated model (\ref{eq:model}) in some specific
limits, for which there are known results. 

The estimated hoppings and Coulomb parameters of the model are summarized in Table \ref{table:parameters}.
\begin{table}
\caption{Estimates of
the Coulomb and hopping parameters parameters for Li$_{0.9}$Mo$_6$O$_{17}$. The on-site Coulomb repulsion, $U$, is obtained from
constrained-DFT calculations  \cite{PopovicPRB06}.  We have assumed: $V =2t$ appropriate for having the material in the 
proximity to a charge order  QCP (see Fig. \ref{fig:phased}). We have assumed a $1/d$ decay of the further 
off-diagonal Coulomb parameters with $d$, the distance between electrons in different sites. 
The Coulomb repulsion energies and hopping amplitudes 
in this table can be identified in the schematic structure of Fig. \ref{fig:struct}. All energies are given in eV. }
\label{table:parameters}
\begin{tabular}{llllllllll}
t  &   t$_\perp$  & t'   &   U  &   V  & V$_\perp$  &  V'  &   V''  &  V''' & V'''' \\
\hline 
0.5  &  -0.024 &  0.036  &  6.4  &  1  &  1  &  0.7  &  0.66  &  0.65 & 0.53\\
\end{tabular}
\end{table}

\section{Understanding the model Hamiltonian}
\label{sec:theory}

In principle, the model (\ref{eq:model}) is
effectively two dimensional consisting of 
zig-zag ladders which are weakly coupled through $t'$ 
 and there is also  
Coulomb repulsion between electrons in different ladders.  The relevance of the interladder hopping interactions is settled by the temperature
scale at which actual experiments are undertaken.  As a first step in the understanding of the full complexity of the model 
we discuss the physics of isolated chains and ladders for which much more theoretical work is available.  
Therefore, we first consider the model (\ref{eq:model}) in different limits assuming independent ladders 
($t'=0$ and 
$V'''=V''''=0$) so we are left only with $t$, $t_\perp$ and the Coulomb repulsion energies: $U$, $V$, $V_\perp$ and $V'$.






\subsection{What is the ground state of the ladder model at 1/4 filling and $U \to \infty$?}

We consider quarter-filled ladders assuming that doubly occupied sites
are forbidden ($U \rightarrow \infty$). This model has been studied numerically using DMRG 
\cite{Vojta} and analytically through weak coupling RG and bosonization 
techniques\cite{Orignac} in different parameter ranges which we now discuss. 
 

\subsubsection{$t_\perp = V_\perp =V'=0$}

The two chains comprising the ladder are completely decoupled
and the model maps onto the $t-V$ model of spinless fermions at half filling.
The ground state is a Luttinger liquid for $V < 2 t$ and an insulator with long range charge order for $ V > 2t$.\cite{NishimotoPRB10}
  Even in the presence of the long range Coulomb interaction decaying as $1/d$,  where $d$ is the distance between electrons
in different sites, it is found that the Wigner state with one electron at every other
site of the lattice is the ground state at one-quarter filling 
 \cite{Hubbard} when $t \rightarrow 0$. 
 This is satisfied in this case indicating that charge ordered states are non-frustrated by the  
long range part of the Coulomb interaction. 
Also recent DMRG calculations on two-leg ladders with $t_\perp=t$ 
show an insulating phase with a charge gap which is interpreted as a result of
dimerization of the rungs with one electron localized on each rung of the ladder\cite{Liu}. 

\subsubsection{$t_\perp = V'= 0, V_\perp \ne 0$}

This ladder model has been studied by bosonisation and RG approaches \cite{Orignac}.
There are two charge modes, total and difference. For $0<V < 2 t $, the difference mode can be gapless. 
The other is gapped in the presence of a small $V_\perp$. This is an homogeneous insulating phase
consistent with DMRG results performed for  $t_{\perp}/t<1$. \cite{Vojta}.
When both $V=V_\perp>2t$ the insulating phase
has zig-zag charge order with wave vector $(\pi,\pi)$. 
The role of $V_\perp$ is crucial in locking
the charge order waves running along the two independent when these are decoupled.
Interestingly,  even at $V=0$ and for any finite $V_\perp$ the system displays a 
charge gap which has an exponential dependence with $V_\perp$ as in the half-filled Hubbard model 
leading to an homogeneous insulator.

\subsubsection{$ V' \neq 0$}

This interaction frustrates charge order
and should produce a metallic state when $V' \sim V \sim V_\perp$,
even when $V \gg t$.\cite{SeoJPSJ06}  In this limit, the model can be mapped
onto a classical Ising model with frustrated interactions which leads to a 
disordered ground state. 
 A quarter-filled one-dimensional extended model \cite{EjimaPRB05} 
with nearest, $V$,  and next-nearest Coulomb interactions, $V'$ has been analyzed using
 DMRG  which shows a metallic state when $V' \sim V/2$.

\subsection{Triplet superconductivity }

Quantum Monte Carlo calculations on a nearly quarter-filled model of weakly coupled chains with onsite 
Coulomb repulsion interaction $U$ only and small $U=2t$ display $f$-wave 
(spin triplet) superconducting tendencies
due to $2k_f$-CDW instabilities. The $f$-wave symmetry is related to the fact that the 
electronic modulation is of about four lattice spacings: $ Q \approx 2k_f \approx \pi/2$ close to quarter-filling\cite{KurokiPRB2004}. 
Such behavior has been confirmed by a random phase approximation (RPA) analysis in 
weakly coupled quarter-filled chains with Coulomb interactions up to third nearest 
neighbors in the presence of interchain Coulomb repulsion. \cite{KurokiJPSJ2005}.
It is found that for  $U=1.7t$, $t_\perp =0.2t$ and under the condition: $V' + V_\perp \approx U/2$, 
triplet  $f$-wave superconductivity wins over the $d$-wave channel.  This is because the presence of $V_\perp$ 
enhances the $2k_f$-CDW instabilities of the isolated chains. 
Triplet superconductivity pairing has also been encountered in weak coupling RG calculations on weakly
coupled chains in the presence of both intrachain and interchain Coulomb repulsion\cite{Nickel}
for moderate values of the interchain Coulomb repulsion.
 
For the two-dimensional quarter-filled extended Hubbard model it is found that introducing frustrating charge interactions on the square lattice
(e.g., along one diagonal) destabilises the stripe charge ordered insulating
phases producing a charge ordered 3-fold symmetric metallic state \cite{Watanabe} at large $V$
and at finite-$U$.  In the limit of $U>>t$ a metallic 'pinball'  liquid state in which
no doubly occupied sites ('pins') can occur \cite{Hotta2006,Canocortes}.  Melting the 3-fold state produces an 'f-wave' 
spin triplet superconducting state. How this particular symmetry emerges can be seen from direct inspection of the 3-fold charge ordering
pattern in real space which consists of placing electrons on the closest sites not coupled via the Coulomb interaction
to a given occupied site,   avoiding the off-site Coulomb repulsion. Thus, the $f$-wave pairing found in the isotropic triangular
lattice is analogous to the $d_{xy}$-wave pairing found on the square lattice \cite{Merino2001} and results from electrons
avoiding the strongest nearest neighbor Coulomb repulsion. 


\subsection{Doping the ladder away from one-quarter filling}

Since Li$_{0.9}$Mo$_6$O$_{17}$ is slightly doped away from one quarter-filling it is
worth considering this situation. Assuming that $t_\perp=V'=0$ and $V_\perp \ne 0$
as above there is a metallic state if the chemical potential is of the order of the gap of the quarter-filled
system\cite{Orignac} based on weak couping RG. This is because a commensurate-incommensurate transition 
occurs since Umklapp processes are suppressed  as: $4k_f \ne |{\bf G}|$ where ${\bf G} $ is a reciprocal vector
of the lattice. On the other hand, superconducting fluctuations in the doped quarter-filled ladder are 
found to be of the singlet $d$-wave type but dominated by the $4k_f$-CDW correlations.  
However, including longer range Coulomb interactions can change this picture and induce $f$-wave
triplet pairing. It remains an open question to understand how the presence of $V'$ 
can influence the superconducting tendencies of the system in ladders doped away from quarter-filling.

\section{Conclusions}

The title quasi-one-dimensional material displays an intriguing competition between insulating, superconducting and
"bad" metallic behavior.   
Besides the mechanism of superconductivity, the nature of the "insulating" phase and the unconventional metallic 
properties are poorly understood. No evidence of a structural transition has been found
accompanying the occurrence of the "insulating" phase. Under pressure the insulating phase 
is suppressed giving way to conventional metallic behavior below a low temperature crossover scale.  

In order to understand these phenomena, we have derived a minimal strongly correlated model for determining the low energy 
electronic properties of Li$_{0.9}$Mo$_6$O$_{17}$. The one-electron part of the Hamiltonian is obtained based on a Slater-Koster
approach and comparison with DFT band structure calculations. The tight-binding Hamiltonian consists 
on three hopping parameters only: $t$, $t_\perp$ and $t'$ which capture the dispersion of the two Mo($d_{xy}$)
bands along the $b$-direction crossing the Fermi energy, the opposite warping of the two
Fermi surface sections, and the weak dispersion of the two bands in the $c$-direction. These are the main 
features found in full band structure calculations.  The real space tight-binding Hamiltonian describes
zig-zag ladders weakly coupled by the small inter-ladder hopping, $t'$. In the lattice model we note that the 
system is close to quarter-filling although the bands are nearly half-filled.  This is a result of the band folding 
associated with the four atoms in the unit cell arising from the zig-zag structure of the ladders. 

A reinterpretation of the physics of Li$_{0.9}$Mo$_6$O$_{17}$ as a nearly quarter-filled system instead of 
a nearly half-filled system is suggested from the model introduced.  Both the onsite $U$ and long range
Coulomb repulsion are found to be relevant since estimated values can be comparable to the bandwidth 
of the material.  Based on these estimates and experiments under pressure we 
suggest that Li$_{0.9}$Mo$_6$O$_{17}$ is close to a charge ordering transition driven by 
the Coulomb repulsion.  Based on the sensitivity of the material to external pressure
we argue that many of the anomalies observed may
arise due to the proximity to a Quantum Critical Point (QCP). 
The "bad" metal behavior is attributed to quantum criticality: a metal with unconventional excitations
arising from the charge fluctuations occurring at all length scales around the QCP. 
Associated with the QCP is the existence of a crossover temperature $T^*$ below which
coherent excitations and Fermi liquid behavior occurs on the metallic side of the 
transition.  This scale is suppressed as the QCP is approached. Quantum fluctuations associated 
with charge order occur in the proximity to  the QCP which can lead to the enhancement 
in the resistivity below $T_m$ leading to a resistivity minimum which has been 
found in other quasi-one-dimensional materials close to CDW instabilities,
such as Per$_2$M(mnt)$_2$ [M=Pt,Au]. \cite{Almeida}

Spin triplet superconductivity can arise in systems in which charge fluctuations
associated with a nearby charge ordered phase dominate.  In general, in  ladder systems explored
close to quarter-filling,  zig-zag charge order correlations are strongly enhanced even by moderate 
interchain Coulomb repulsion, $V_\perp$, as previous works have shown. 
In Li$_{0.9}$Mo$_6$O$_{17}$, the fact that $V_\perp = V$ and the presence of next-nearest neighbors
Coulomb interaction and longer range interactions  along the chains due to the zig-zag structure can
favor spin triplet superconductivity.  

Experiments probing the existence of large charge fluctuations near the QCP are desirable
in order to understand the enhancement of the resistivity with decreasing temperature.  X-ray diffraction
find no structural changes associated with the upturn of the resistivity at $T_m$ suggesting that a
purely electronic mechanism plays a major role.  From the $T$-dependence of the 
$1/T_1T$-NMR spectra one could find whether anomalous broadening or splitting of spectral lines\cite{Hiraki} occurs
when lowering the temperature below $T_m$ in analogy to 
TMTTF$_2$AsF$_6$ 
(Ref. \onlinecite{Chow}) ,
in which the charge ordering transition detected by NMR is not accompanied by a change of the structure. 
This  kind of experiments could clarify the role played by charge ordering phenomena driven by the 
long range Coulomb repulsion as proposed here.


 \begin{acknowledgments}
We thank Nigel Hussey for stimulating our interest in this material and for many helpful discussions. 
J. M. thanks J. V. \'Alvarez for helpful insights.  
R.M. received financial support from an
  Australian Research Council Discovery Project grant (DP0877875).
 J. M. acknowledges financial support from the International Collaboration Agreement 
 from Australian Research Council Discovery Project (DP10932249)
[Chief Investigator: B.J. Powell] and
 Ministerio de Econom\'ia y Competitividad in Spain (MAT2011-22491).
  
 \end{acknowledgments}


\begin{thebibliography}{39}
\bibitem{1DHall}
In a quasi-one-dimensional system band structure effects leads to a modified Drude expression\cite{Yakovenko99}: $R_H={\beta \over n|e|}$, where
$\beta=-{k_F \over v_F}{\partial^2 \epsilon(k_F) \over \partial k_b^2 }$, with $v_F=|{\partial \epsilon(k_F) \over \partial k_b}|$
and $k_F$ the Fermi wavevector. In a quasi-one-dimensional crystal with dispersion: $\epsilon(k_b)=-2t\cos{(k_bb/2)}$, ($t>0$) 
the simple Drude expression is modified 
to: $R_H=-\frac{1}{n|e|}\frac{k_F b/2}{\tan (k_F b/2)}$, where $b/2$ is the lattice constant along 
the chain direction.\cite{Cooper,Virosztek}  For $k_F = {\pi \over 2b}$ relevant to Li$_{0.9}$Mo$_6$O$_{17}$, we have $R_H=-{\pi \over 4} {1 \over n|e|}<0$.
\end{thebibliography}
\end{document}